\documentclass[aps,prd,preprint,floatfix,nofootinbib]{revtex4}
\usepackage{graphicx}
\newcommand{\be}{\begin{equation}}
\newcommand{\ee}{\end{equation}}
\newcommand{\ba}{\begin{array}}
\newcommand{\ea}{\end{array}}
\newcommand{\bea}{\begin{eqnarray}}
\newcommand{\eea}{\end{eqnarray}}

\newcommand{\tr}{^{\rm tr}}

\newcommand{\lt}{\left}
\newcommand{\rt}{\right}

\newcommand{\SU}{{\rm SU}}

\newcommand{\chibar}{{\overline\chi}}
\newcommand{\jbar}{{\overline\jmath}}
\newcommand{\lesim}{${\lower 2pt\hbox{$\scriptstyle
<$}\atop\raise 4pt\hbox{$\scriptstyle\sim$}}$} 
\newcommand{\grsim}{${\lower2pt\hbox{$\scriptstyle >$} \atop\raise4pt\hbox 
{$\scriptstyle\sim$}}$}

\newcommand{\Go}{{\cal G}}
\begin{document}
 
\title{Cold, dense matter via the lattice NJL model
\footnote{Talk presented at the Workshop on QCD in Extreme
Environments, Argonne National Laboratory, IL, USA,
29$^{\hspace{0.1em}t h}$ June to 3$^{\hspace{0.1em}r d}$ July, 2004.}}

\author{David~N.~Walters}
\affiliation{Theoretical Physics Group, Department of Physics and
Astronomy,\\ University of Manchester, Manchester M13 9PL, United Kingdom}
  
\begin{abstract}

We simulate the lattice Nambu--Jona-Lasinio (NJL) model in
$3+1$-dimensions at non-zero baryon chemical potential ($\mu$) and zero
temperature ($T$) and treat the results as phenomenologically relevant for
cold, dense quark matter.
Measurements of the chiral condensate indicate a crossover in the
thermodynamic limit, whilst at high chemical potential and zero
temperature we observe a non-zero diquark condensate and a gap in
the fermion dispersion relation, which together provide evidence for BCS
superfluidity. In particular, the size of gap is found to be
approximately $15\%$ the value of the vacuum fermion mass and roughly
independent of $\mu$ in the chirally restored phase.
\end{abstract}

\maketitle

\section{Introduction} 

Whilst many of the talks at this workshop have been concerned with
QCD at extreme temperatures, the ``extreme environment'' in the work
presented here is that of cold, dense matter, where QCD is believed to
exhibit colour superconductivity via the condensation of diquarks
analogous to the Cooper pairs of BCS superconductors (for recent
reviews on colour superconductivity see
e.g.~\cite{Rajagopal:2000wf,Rischke:2003mt}).
The BCS mechanism in a superconductor is a subtle one, since the
fundamental interaction between electrons is repulsive and
the net attractive interaction due to phonon exchange persists only at
extremely low temperatures. In QCD, however, the fundamental
interaction between quarks is attractive in the anti-triplet channel,
such that diquark condensation should be far more robust against
thermal fluctuations. In recent years, studies of four-Fermi models
with QCD instanton motivated interactions have suggested that
the BCS gap in QCD could be as large as
50-100MeV~\cite{Berges:1998rc}.  This means that a colour superconducting
phase could be relevant to the physics of compact stars
where the typical temperature is only ${\cal
O}$(1MeV). Analytic studies have shown that the ground-state of
$2+1$ flavour QCD at asymptotically high density is the colour-flavour-locked
phase~\cite{Alford:1998mk}. Due to the persistence of the sign
problem in lattice QCD with $\mu\neq0$, however, to investigate the nature of QCD at more moderate densities
one must to resort to studying model field theories. The
non-perturbative treatment of such models can potentially provide a robust
method to study colour superconducting matter at the intermediate
densities relevant to the physics of compact stars. 

QCD with an $\SU(2)$ gauge group is one interesting model which can be
studied non-perturbatively as it does not suffer from a sign
problem~\cite{Hands:2000ei}. Lattice studies in recent years have
shown that two colour QCD exhibits superfluidity  via
Bose-Einstein condensation in the dense phase in a manner analogous to
He$^4$,  due to the bosonic nature of the two-quark
baryons~\cite{Aloisio:2000rb,Kogut:2001na,Hands:2001ee}.  
In order to observe a BCS-style scenario, however, it seems that
one must resort to studying purely fermionic field theories such as the
Nambu--Jona-Lasinio (NJL) model~\cite{Nambu:1961tp}. This model has no
gauge degrees of freedom and all orders of gluon exchange are
approximated by a point-like four-fermion interaction.
The original motivation for the formulation of this model was
that it observed the same global symmetries as strongly
interacting matter and its vacuum structure exhibited chiral symmetry
breaking in a manner directly analogous to BCS
superconductivity. Therefore, whilst it cannot teach us anything about the
mechanism of (de)confinement or the interaction of charged diquarks, if one
constricts oneself to studying its global symmetries and the pattern
of their breaking, it can be an ideal model with which to study colour 
superconductivity in a relativistic quantum field theory. 

This talk contains results from a numerical study of the high $\mu$,
low $T$ phase of the $3+1$-dimensional NJL model, published
in~\cite{Hands:2004uv}, with the aim of showing that the ground-state
of this model is that of a conventional BCS superfluid formed in a
manner analogous to superfluid He$^3$. After describing the
formulation of the model on the lattice in Sec.~\ref{sec:model}, we
present measurements of the chiral and diquark condensates and discuss
the zero temperature and infinite volume limits in Sec.~\ref{sec:phase}.
Finally, we present a direct measurement of the BCS gap in
Sec.~\ref{sec:gap} and show that it is consistent with model predictions.

\section{The lattice NJL model}
\label{sec:model}

The action of the lattice NJL model, with the lattice spacing $a\to1$,
can be written as\footnote{Here, the bosonic part is corrected from
that of the 
equivalent expression in~\cite{Hands:2004uv} by a factor of 2, which
appeared therein due to a typographical error.}
\begin{equation}
S=S_{f e r m}+S_{b o s}=\frac{1}{2}\sum_{x y}\Psi_x\tr{\cal A}_{x y}\Psi_y+
\frac{1}{g^2}\sum_{\tilde x}{\rm t r}\Phi^\dagger_{\tilde x}\Phi_{\tilde x},
\label{eq:Slatt}
\end{equation}
where the bispinor $\Psi$ defined on lattice sites $x$ is written in
terms of independent isospinors via
$\Psi\tr\equiv\lt(\chibar,\chi\tr\rt)$ and the auxiliary scalar and
pseudoscalar fields defined on dual lattice sites $\tilde x$ are
introduced via the $2\times2$ matrix $\Phi\equiv\sigma+i\vec\pi.\vec\tau$.
The kinetic operator ${\cal A}$ written in the Nambu-Gor'kov basis is given
by
\be
{\cal A}=\lt(\ba{cc}\jbar\tau_2&M\\-M\tr&j\tau_2\ea\rt),
\ee
where $\jbar$ and $j$ are U(1)$_B$ symmetry breaking terms, which fix the
direction of symmetry breaking to allow the
measurement of a diquark condensate on a finite volume lattice in
analogy with the bare quark mass in the measurement of the
chiral condensate. In practice, we use $\jbar$ and $j$ real, positive
and equal, and what we refer to as $j$ from this point on is actually the sum
of these two terms. Also, simulations are performed in the ``partially
quenched'' approximation, with $j$ being zero during the generation of
our background field configurations, and being made non-zero only
during the measurement of the observables. This makes the simulations
far less computationally expensive, and it should be noted that in
simulations of the model in $2+1$-dimensions, there was no discernible
difference between measurements made using this method and those made
via full simulation~\cite{Hands:2001aq}. Finally, $M$ is the standard fermion kinetic operator 
\begin{eqnarray}
M_{x y}^{p q}&=&{\frac12}\delta^{p q}\left[
e^\mu\delta_{y x+\hat0}-e^{-\mu}\delta_{y x-\hat0}+\sum_{\nu=1,2,3}\eta_\nu(x)
(\delta_{y x+\hat\nu}-\delta_{y x-\hat\nu})+2m\delta_{x y}\right]\nonumber\\
&+&{\frac{1}{16}}\delta_{x y}\sum_{\langle\tilde x,x\rangle}\lt(\sigma(\tilde
x)\delta^{p q}+i\varepsilon(x)\vec\pi(\tilde x).\vec\tau^{p q}\rt),
\label{eq:M}
\end{eqnarray}
with the parameters being the bare fermion mass $m$, coupling $g^2$ 
and baryon chemical potential $\mu$. The symbol $\langle\tilde
x,x\rangle$ denotes the set of 16 dual sites adjacent to $x$. 
The Pauli matrices acting on isospin indices $p,q$ are normalised so that
${\rm t r}\tau_i\tau_j=2\delta_{i j}$. The phase factors
$\eta_\nu(x)=(-1)^{x_0+\cdots+x_{\nu-1}}$ and
$\varepsilon(x)=(-1)^{x_0+x_1+x_2+x_3}$ ensure that fermions with the correct
Lorentz covariance properties emerge in the continuum limit, and that in the
limit $m\to0$ the action (\ref{eq:Slatt}) has a global
SU(2)$_L\otimes$SU(2)$_R$ invariance. In addition, in the limit that
$j\to0$, the action has a U(1)$_B$ invariance under baryon number rotations.

Dimensional analysis shows that the coupling $g^2$ has mass dimension
$-2$, which reflects the fact that in $3+1$-dimensions, four-fermion
models have no non-trivial continuum
limit. We deal with this by fitting our model to low energy, vacuum
phenomenology and extracting the relevant bare
parameters~\cite{Klevansky:1992qe}. In particular, we calculate
dimensionless ratios of the constituent fermion mass and the mass and
decay rate of the pion. Setting these to $400$MeV, $138$MeV and
$93$MeV respectively fixes our bare parameters as $a m=0.006$ and
$a^2/g^2=0.495$. For details, see~\cite{Hands:2004uv}.

\section{Phase structure and the zero temperature limit} 
\label{sec:phase}

Self-consistent treatment of the NJL model shows that for sufficiently
strong coupling, the  approximate $\SU(2)_L\otimes\SU(2)_R$ chiral
symmetry is  spontaneously broken in the vacuum to $\SU(2)_V$,  leading
to a dynamically generated  quark mass $\Sigma\gg m$, and 3
degenerate pseudo-Goldstone modes identified with the pions.  In the
presence of a baryon chemical potential $\mu\neq0$ the symmetry is
approximately restored as $\mu$ is increased through some onset scale
$\mu_o\sim\Sigma$, with the order of the transition being sensitive to the parameters employed~\cite{Klevansky:1992qe}.

We determine the nature of this transition in the regime set by our
phenomenological parameter choice by studying the order parameter of
chiral symmetry breaking, the chiral condensate $\lt<\chibar\chi\rt>$,
defined by 
\be
\langle\bar\chi\chi\rangle={\frac1V}{\frac{\partial\ln{\cal Z}}{\partial m}}=
{\frac1{2V}}\left\langle\mbox{tr}\lt(
\ba{c c}&\openone_{2\times2}\\-\openone_{2\times2}&\ea\rt)
{\cal A}^{-1}\right\rangle, 
\label{eq:chibarchi}
\ee
where ${\cal Z}\equiv\int{\rm d}\chi{\rm d}\bar{\chi}{\rm d}\Phi e^{-S}$ is
the partition function associated with (\ref{eq:Slatt}). 
$\lt<\chibar\chi\rt>$ is measured for a range of $\mu$ over 500
equilibrated  hybrid Monte Carlo trajectories of length 1.0 on lattice
volumes $V=L_s^3\times L_t=12^4$, $16^4$ and $20^4$ with the trace
taken via 5 stochastic estimations made on every other
configuration. 
\begin{figure}[h]
\centering
\includegraphics[width=10cm]{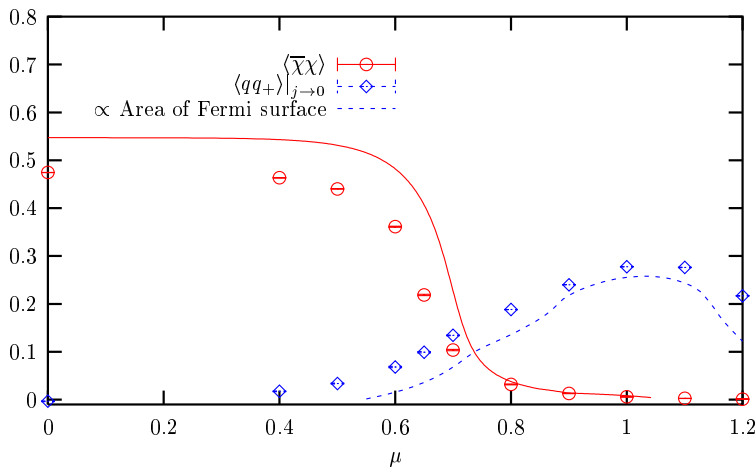}
\caption{Chiral and diquark condensates as functions of chemical potential.}
\label{fig:EofS}
\end{figure}
The data were then extrapolated linearly in $V^{-1}$ to the infinite
volume limit. The data are presented as the circular points in
Fig.~\ref{fig:EofS}. As expected, we see that chiral symmetry is
broken  at zero chemical potential, as signified by a large chiral
condensate, and is approximately restored  after the condensate passes
through a broad crossover and becomes approximately zero in the
high $\mu$ phase. We can also calculate $\lt<\chibar\chi\rt>$
analytically via a self-consistency equation (the gap equation) in the
limit that the number of fermion species (which we rather arbitrarily
refer to as the number of colours $N_c$) is taken to infinity. This
is plotted as the solid curve and can be seen to qualitatively agree
with the lattice data, in which the ${\cal O}(1/N_c)$ corrections are
$\approx15\%$.

\begin{figure}[h]
\centering
\includegraphics[width=11cm]{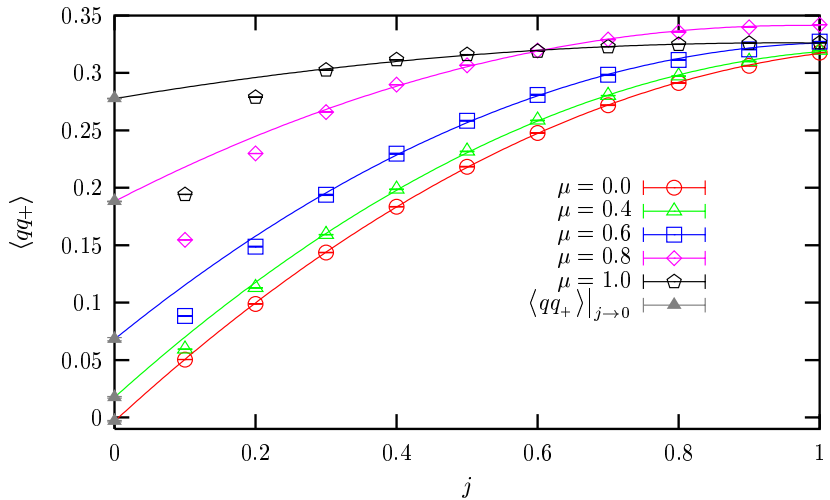}
\caption{Diquark condensate as a function of $j$ in the zero
temperature limit for various $\mu$.} 
\label{fig:qqvsj}
\end{figure}

In order to explore the possibility of a U(1)$_B$-violating BCS phase
at high $\mu$ we study the relevant order parameter, the diquark
condensate $\lt<q q_+\rt>$, which in analogy with the chiral
condensate is defined by
\be
\lt<q q_+\rt>=\frac{1}{V}\frac{\partial\ln{\cal Z}}{\partial j}
=\frac1{4V}\lt<{\rm tr}\lt(\ba{c c}\tau_2&\\&\tau_2\ea\rt){\cal
A}^{-1}\rt>.
\ee
Here, the positive subscript highlights the fact that $\lt<q q_+\rt>$
is the sum of contributions from both $\chi\chi$ and
$\bar{\chi}\bar{\chi}$ states. This expectation value is measured for
ten values of $j$ between 0.1 and 1.0 on the same volumes and with the
same parameters with which we measure $\lt<\chibar\chi\rt>$. Unlike
$\lt<\chibar\chi\rt>$, however, the data are found not to scale with
inverse volume, but instead appear to scale linearly with inverse
temporal extent. Accordingly, we extrapolate our data to
$L_t^{-1}\to0$ corresponding to the zero temperature limit. Some of
these data are presented in Fig.~\ref{fig:qqvsj}.
In contrast with the measurement of $\lt<\chibar\chi\rt>$, in which the
bare quark mass is left non-zero as the chiral symmetry of QCD is
believed to only ever be approximately conserved, we are required to
extrapolate the data for $\lt<q q_+\rt>$ to the $j\to0$ limit, in which
the phenomenologically exact U(1) symmetry related to the conservation
of baryon number is restored. By empirically fitting a quadratic
curve through the $\mu=0$ data, one finds that as $j$ goes to zero so
does $\lt<q q_+\rt>$, implying that there is no condensation  in the vacuum
as expected. For values of $\mu$ in the chirally restored phase,
however, the picture is not quite so clear. It appears that the
data correspond to a non-zero diquark condensate, but one can only fit a
curve though a subset of the data such that one is required to
disregard data with $j\le0.2$. Whilst the resulting curves appear to
fit the remaining data well, it is important to be able to justify this
omission, especially as the data omitted are those closest to
limit which one is attempting to reach. 

\begin{figure}[h]
\centering
\includegraphics[width=11cm]{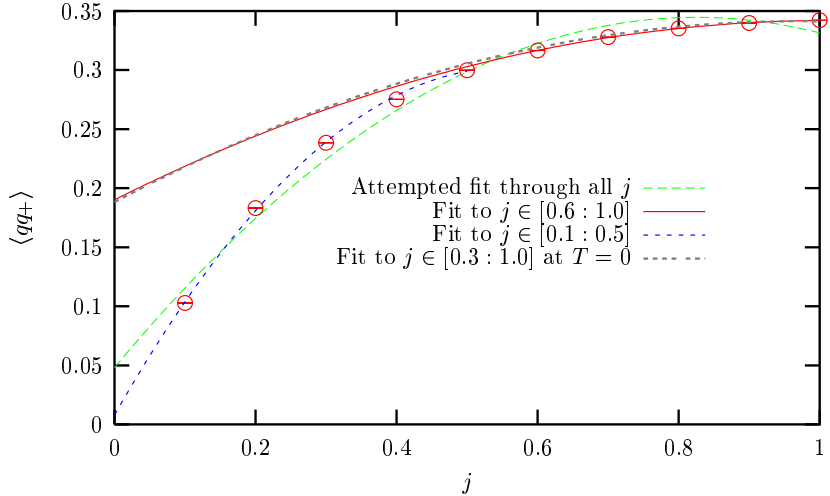}
\caption{Diquark condensate as a function of $j$ at $\mu=0.8$ on
a $30^3\times12$ lattice.}
\label{fig:lt_12}
\end{figure}

The most obvious possibility
is that the condensate is suppressed due to finite volume effects,
since diquark condensation is associated with the breaking of a global
symmetry; as the symmetry breaking source is reduced to zero, the
correlation length $\xi$ of the fluctuations of the order parameter
should diverge and become comparable to the size of the lattice.
\begin{figure}[h]
\centering
\includegraphics[width=11cm]{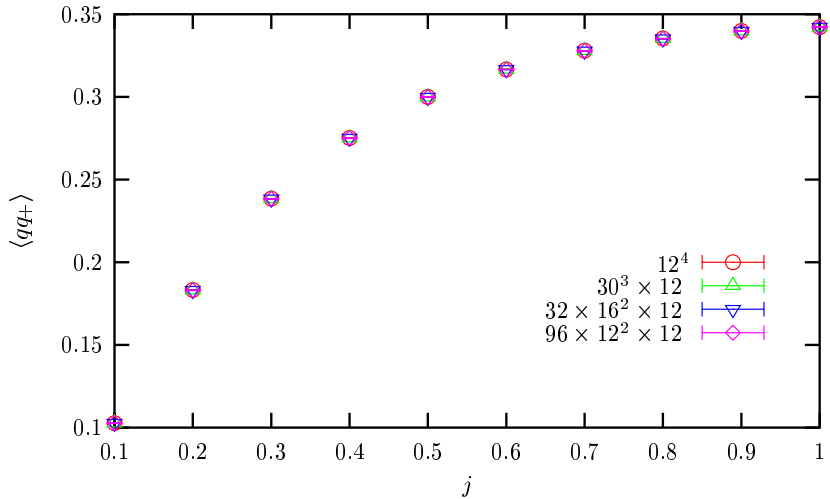}
\caption{Diquark condensate vs. $j$ at $\mu=0.8$ on
$L_t=12$ lattices with various spacial volumes.}
\label{fig:finitev}
\end{figure}
Figure~\ref{fig:finitev}, however, shows some results from an extensive
finite volume study, which illustrates that $\lt<q q_+\rt>$ shows
little or no change for any value of $j$, even when the spatial volume
is altered by a factor of $2.5^3\approx16$. 
It seems impossible, then,
that this is the source of the suppression of the condensate at low
$j$. Another possibility is that this suppression is due to having
poor control over the extrapolation to zero temperature.
\begin{figure}[ht]
\centering
\includegraphics[width=11cm]{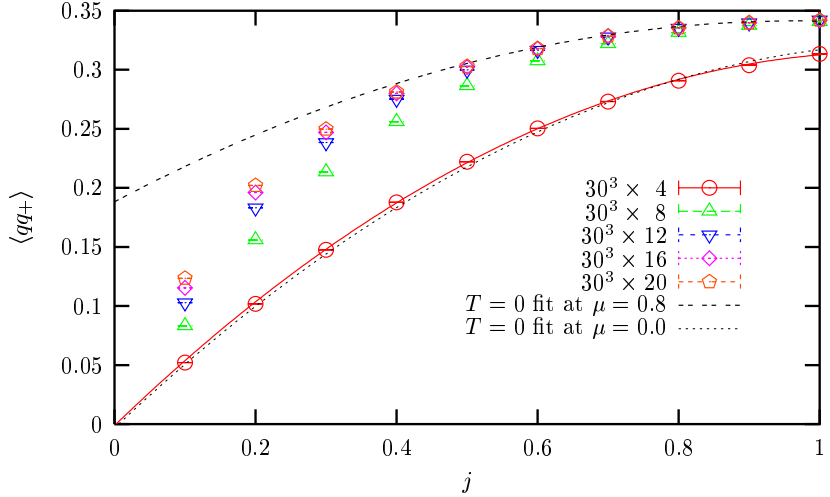}
\caption{Diquark condensate as a function of $j$ at $\mu=0.8$ at
various temperatures.}
\label{fig:Tneq0}
\end{figure}
Figure~\ref{fig:Tneq0} shows the results of a study of the model at
$\mu=0.8$ and with various inverse temporal extents, corresponding to
various non-zero temperatures, as well as the curves fitted to
$j\in[0.3,1.0]$ of our $\mu=0.0$ and $\mu=0.8$ data after the $T\to0$
extrapolation. At the highest temperature studied, for which $L_t=4$,
if one performs a quadratic extrapolation through the data one observes that
$\lt<q q_+\rt>=0$ in the $j\to0$ limit and that the curve closely
resembles the $\mu=0.0$ curve at zero temperature. This suggests that
although we are within the chirally restored phase, the temperature
with $L_t=4$  is above the critical temperature for superfluidity for
all $j$. As the temperature is decreased (i.e.~as the $L_t$ is
increased to 8) the value of $\lt<q q_+\rt>$ at $j=1.0$ immediately
approaches the curve fitted to our zero temperature results, whilst
for lower values of $j$ this only occurs at even lower
temperatures. This can be understood if one considers the fact that the
effect of $j$ is to make the condensate robust not only to finite
volume effects, but also to other perturbations such as thermal
fluctuations. For a given value of $j$, therefore, there should be
some pseudo-critical temperature $T_c(j)$ below which the data can be
said to be in the superfluid phase, and which in the thermodynamic
limit should increase monotonically with $j$. This idea is well
illustrated by the data measured on the $30^3\times12$ lattice, shown
in Fig.~\ref{fig:lt_12}. 
An attempted fit to a single quadratic through all values of $j$ is of
very poor quality, whilst by choosing to split the data at some
suitable point and fitting the two regions separately, two very
reasonable fits can be obtained. In particular, a fit to the data with
$j\le0.5$ is consistent with $\lt.\lt<q q_+\rt>\rt|_{j\to0}=0$
suggesting that for these values of $j$, $L_t=12$ is above the
pseudo-critical temperature. The fit to the data with $j\ge0.6$,
however, is consistent with $\lt.\lt<q q_+\rt>\rt|_{j\to0}\neq0$ and
agrees almost exactly with the fit to data from the $12^4$,
$16^4$ and $20^4$ lattices extrapolated to $T\to0$ and then
$j\in[0.3,1.0]\to0$, suggesting that this curve represents the true
zero temperature limit and justifies our discarding of data with
$j\le0.2$.  Whilst a linear $T\to0$ extrapolation is sufficient to
reach this curve for $0.3\le j\le0.5$, the condensate at $j<0.3$ must be
suppressed too much for such an extrapolation to be sufficient,
i.e. the temperature of the lattice on our three lattice volumes must
be too high compared with $T_c(j<0.3)$.

Finally, now that we can trust our $T\to0$ and $j\to0$ extrapolations,
let us look back at Fig.~\ref{fig:EofS}, where the diquark condensate
is plotted with the chiral condensate as a function of $\mu$. Although
there is clearly a transition from a phase with no diquark condensation
to one in which the diquark condensate has a magnitude similar to that
of the vacuum chiral condensate, this transition is far less
pronounced than in the chiral case. $\lt<q q_+\rt>$ increases approximately
as $\mu^2$, but eventually saturates as $\mu$ approaches 1.0 and even
decreases past $\mu\sim1.1$. This behaviour is 
directly related to the geometry of the Fermi surface for a system defined on a
cubic lattice, the area of which we can calculate in the large-$N_c$
free fermion limit and is plotted as the dashed curve. 
In the continuum, therefore, $\lt<q q_+\rt>$ should continue to rise as
$\mu^2$.

\section{Measurement of the gap}
\label{sec:gap}

In the previous section, we presented evidence for superfluidity in
the form of a non-zero diquark condensate at high chemical
potential. Here we present more direct evidence for a superfluid phase
by mapping out the fermion dispersion relation (i.e. the energy $E$ as
a function of momentum $k$) and observing a BCS energy gap $\Delta$
about the Fermi momentum. One advantage of this order parameter over
the diquark condensate is that it is directly related to a macroscopic
property of the superfluid, the critical temperature
$T_c$~\cite{Schmitt:2002sc}. Also, being a global order parameter, in
principle it
would be possible to measure in a gauge theory such as QCD, where
according to Elitzur's theorem, one cannot write down a local order
parameter such as $\lt<q q_+\rt>$ in a gauge invariant
way~\cite{Elitzur:1975im}. 

In order to extract the dispersion relation we define the
time-slice propagator 
\be
\Go(\vec k;t)=\sum_{\vec x}\Go(\vec0,0;\vec x,t)e^{-i \vec k.\vec x},
\label{eq:timeslice}
\ee
where
\be
\Go(x;y)={\cal A}^{-1}_{x y}
=\lt(\ba{c c}A_{x y}&N_{x y}\\\bar{N}_{x y}&\bar{A}_{x y}\ea\rt)
\ee
is the Gor'kov propagator.
As in the original BCS theory~\cite{Bardeen:1957mv}, the
fermionic degrees of freedom can be viewed as
quasi-particles  with energy $E$ relative to the system's Fermi
energy $E_F$. In the limit that $j\to0$,
the propagation of these  quasi-particles is described by the ``normal''
$\lt<q(0)\overline q(x)\rt>$ and $\lt<\overline q(0)q(x)\rt>$ 
parts of (\ref{eq:timeslice}), i.e. those that are off-diagonal in the 
Nambu-Gor'kov space and related to $M^{-1}$. 
If the Fermi surface is unstable with respect to a BCS
ground-state,  the quasi-particles nearest to $E_F$
undergo particle-hole mixing and a gap appears in the energy
spectrum. The propagation of these mixed states is generated by the
diagonal, or ``anomalous'' $\lt<q(0)q(x)\rt>$ and $\lt<\overline
q(0)\overline q(x)\rt>$ parts of (\ref{eq:timeslice}). By a
combination of symmetry constraints and empirical observations we note
that the complex matrix $\Go(\vec k,t)$ contains only two independent
parts, one in $\bar{N}_{x y}$ and one in $A_{x y}$, which from hereon shall
be referred to simply as the normal and anomalous propagators and
written as $N(k,t)$ and $A(k,t)$ respectively.

We measure $N(k,t)$ and $A(k,t)$ at $\mu=0.8$ on lattices with
$L_x\times L_y\times L_z=96\times12\times12$ and $L_t=16$, $20$ and
$24$ using standard lattice techniques and extrapolate linearly in
$L_t^{-1}$ to zero temperature. This choice means that by choosing
$\vec k=(k,0,0)$ in (\ref{eq:timeslice}) we can study 25 independent
momentum modes in the $x$ direction between 0 and $\pi/{2 a}$. We may
then extract the energy by fitting them to
\be\ba{c l l}
N(k,t)=A e^{-E t}+Be^{-E(L_t-t)}&{\rm if}&t={\rm odd}\\
N(k,t)=0&{\rm if}&t={\rm even}
\label{eq:Nfit}
\ea\ee
and
\be\ba{c l l}
A(k,t)=C(e^{-E t}-e^{-E(L_t-t)})&{\rm if}&t={\rm even}\\
A(k,t)=0&{\rm if}&t={\rm odd},
\label{eq:Afit}
\ea\ee
where $A$, $B$ and $C$ are kept as free
parameters, as is the energy $E$, which as expected is found to be the same 
from both (\ref{eq:Nfit}) and (\ref{eq:Afit}). These parameters are
then, in turn,
extrapolated to $j\to0$. Quadratic polynomial curves are fitted to
the coefficients $A(k)$, $B(k)$ and $C(k)$, whilst the energy $E(k)$
is fitted with a straight line. As with the extrapolation of $\lt<q
q_+\rt>$ in the previous section, the
extrapolations appear to smoothly fit the data except for at low $j$,
where the discrepancy we have attributed to non-zero temperature
persists. Again, for the purpose of the extrapolations, we believe we
are justified in ignoring points with $j<0.3$.  

\begin{figure}[ht]
\centering
\includegraphics[width=10cm]{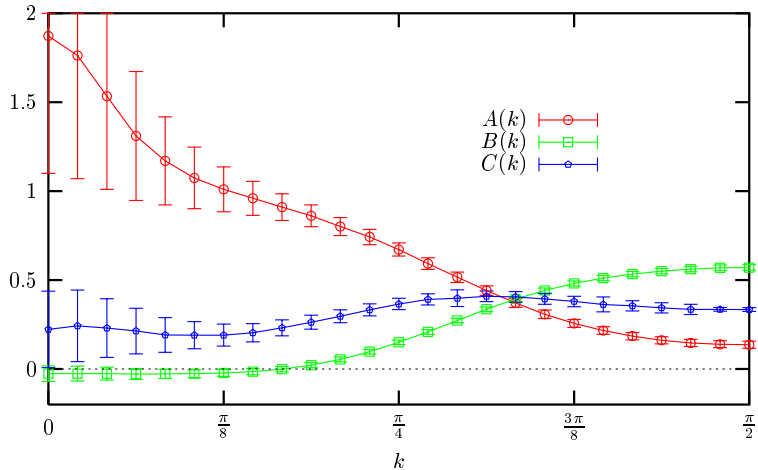}
\caption{Propagator coefficients $A$, $B$ and $C$ at $\mu=0.8$ in the
zero temperature and source limits.} 
\label{fig:ABCvsk}
\end{figure}
Figure~\ref{fig:ABCvsk} shows the coefficients of the normal and
anomalous propagators, after the extrapolations to zero temperature
and diquark source, plotted as functions of momentum $k$. Since the
normal propagator was chosen from the $\bar{N}_{x y}$ part of $\Go$,
the forward moving signal proportional to the coefficient $A(k)$ relates
to the propagation of holes in the Fermi sea, whilst the backward
moving signal proportional to $B(k)$ relates to the propagation of
particle excitations above the Fermi surface. Accordingly, we see that
the low momentum excitations near the centre of the Fermi sphere are
dominated by hole degrees of freedom, whilst the high momentum
excitations are dominated by particles. To excitations at the Fermi
surface, the propagation of fermions forward and backward in time
should be equal, such that the point where $A(k)$ and $B(k)$ cross
allows us to define the Fermi momentum $k_F$. The coefficient
$C(k)\sim0$ at low momentum, but becomes non-zero in a broad peak
about the position of $k_F$. This vanishing of the anomalous
propagator $A(k,t)$, even in the limit that $j\to0$ is a signal of
particle-hole mixing in the presence of a BCS gap.

\begin{figure}[ht]
\centering
\includegraphics[height=6cm]{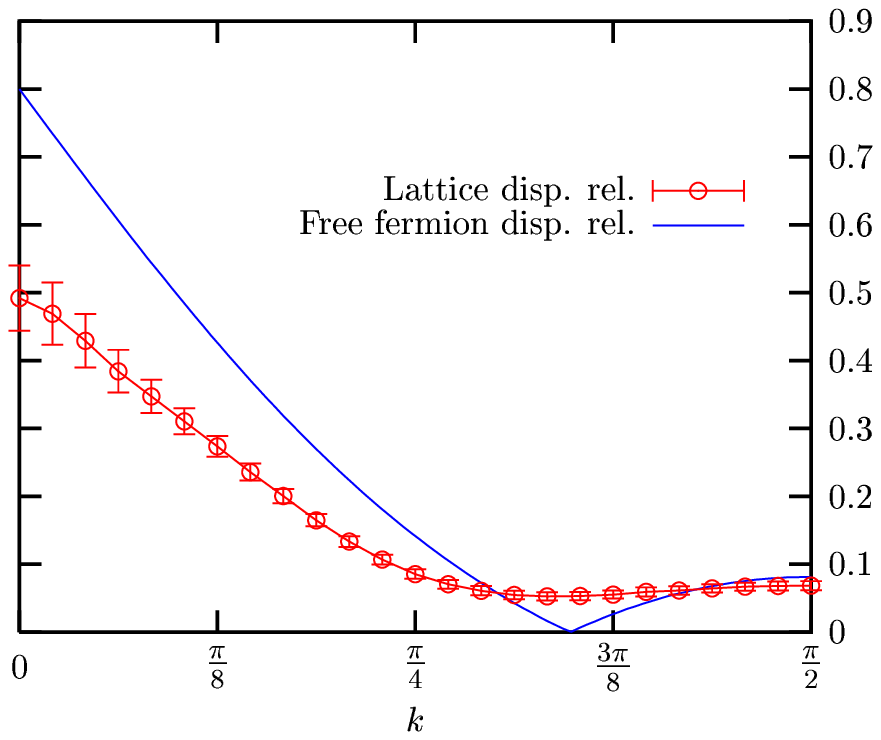}
\includegraphics[height=6cm]{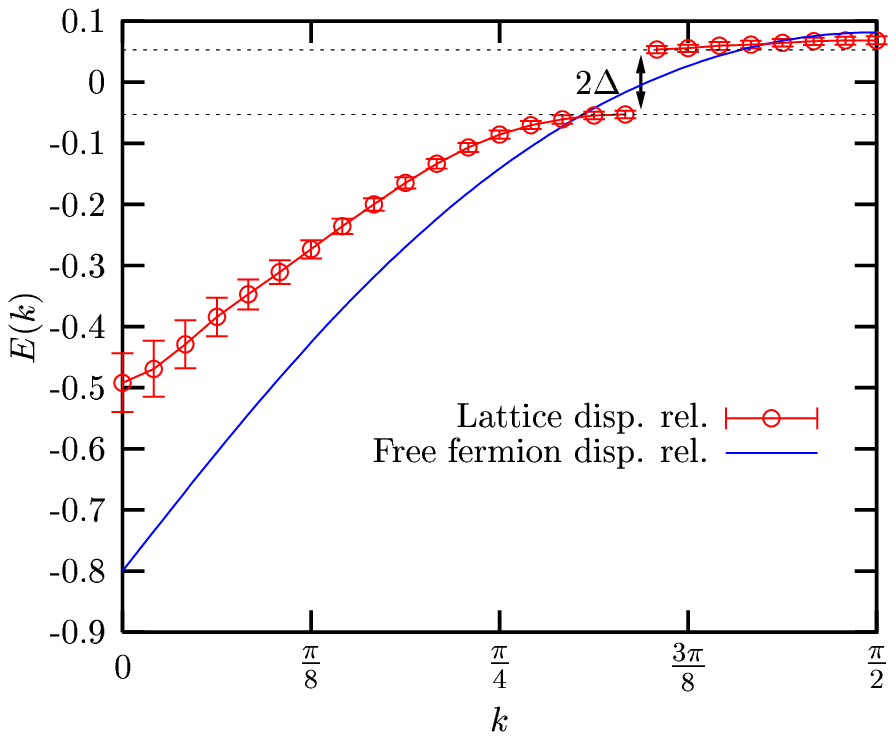}
\caption{The lattice dispersion relation and typical free fermion
dispersion relation at $\mu=0.8$. In the right-hand panel the hole
branch is plotted as negative.}
\label{fig:Evsk}
\end{figure}

For more direct evidence, let us look at the left panel of
Fig.~\ref{fig:Evsk}, which shows the dispersion relation at $\mu=0.8$
and again extrapolated to $T\to0$ and $j\in[0.3,1.0]\to0$. 
The solid curve shows the dispersion relation for free massless
staggered fermions, which has two distinct branches, the hole branch
where $E(k)$ decreases with $k$ corresponding to excitations below
$E_F$ and the particle branch where $E(k)$ increases with $k$
corresponding to excitations above $E_F$. In contrast to this, the
dispersion relation from our lattice data shows no  discontinuity
between the two branches, which is another sign of particle-hole
mixing. More importantly, at no time does the curve pass through
$E=0$ as there is a distinct gap between this point and the minimum;
this is the BCS gap $\Delta=0.053(6)$. This can be seen in a more
familiar light if we plot the hole branch as negative, as illustrated
in the right-hand panel of Fig.~\ref{fig:Evsk}. This makes the free
fermion dispersion relation a smooth continuous curve, as one would
expect, whilst for the NJL dispersion relation this introduces a
discontinuity at $k_F$, which gives a clear illustration of the BCS
gap. In order to present the value of the gap as a dimensionless
ratio, we also measure the fermion mass at $\mu=0$ from the vacuum
dispersion relation which gives us
\be
\frac{\Delta(\mu=0.8)}{\Sigma(\mu=0.0)}=0.15(2).
\ee
Assuming a fermion mass of 400MeV, this implies that
$\Delta(\mu=0.8)\approx60{\rm MeV}$, consistent with the analytic
predictions of~\cite{Berges:1998rc,Nardulli:2002ma}.

Finally, in order to investigate the $\mu$ dependence of $\Delta$, we
determine dispersion relations for a range of chemical potentials in
$0.50\le\mu\le0.85$, this time using data from $L_t=16$ and 20
only. Whilst this means that the extrapolation to $T=0$ is no longer an
overdetermined problem and requires one to estimate the error, it is
worth noting that the resulting dispersion relation at $\mu=0.8$
agrees with that presented in Fig.~\ref{fig:Evsk} for all $k$.
\begin{figure}[ht]
\centering
\includegraphics[width=8cm]{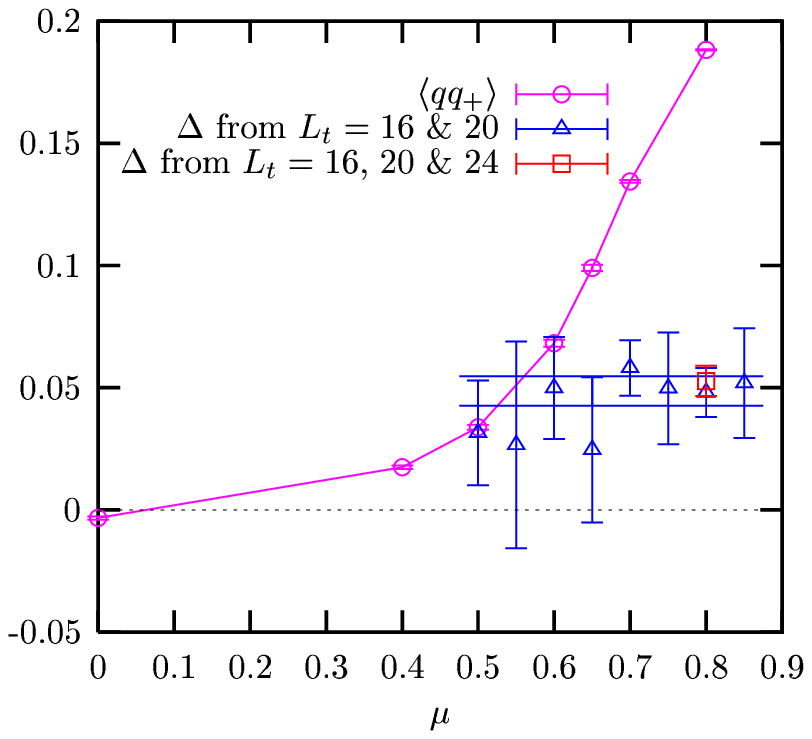}
\caption{The gap $\Delta$ as a function of $\mu$ compared with $\lt<q q_+\rt>$}
\label{fig:Evsmu}
\end{figure}
By calculating $E(k)$ and extracting the minimum, we are able to plot
$\Delta(\mu)$ in Fig.~\ref{fig:Evsmu}, where it is compared with the
diquark condensate $\lt<q q_+\rt>$. Whilst the condensate rises with
$\mu$, the gap shows no evidence of any $\mu$ dependence in the
chirally restored phase and a least-squares fit to $\Delta=constant$
has a $\chi^2$ of only 0.33 per degree of freedom. In combination
with Fig.~\ref{fig:EofS}, this supports the simple-minded picture in which
only quark pairs within a shell about $E_F$ of thickness $2\Delta$,
independent of $\mu$, contribute to diquark condensation, resulting in
a condensate $\lt<q q_+\rt>\propto\Delta\mu^2$.

\section{Summary}

In this talk we have seen evidence, first presented
in~\cite{Hands:2004uv}, for a BCS superfluid phase at high
$\mu$ and low $T$ in the $3+1d$ lattice
NJL model in the form of a non-vanishing local order parameter and,
for the first time in the systematic study of a relativistic field theory,
a direct observation of a BCS gap.  
Given that the model shares its global symmetries with the real world,
we believe that we are justified in treating this as phenomenological
evidence for the existence of a similar colour superconducting phase
in full QCD. 

\section*{ACKNOWLEDGEMENTS}

The author would like to thank Don Sinclair and the staff at the Argonne
National Laboratory for organising the workshop and for
their financial support and hospitality.  
This work was performed in collaboration with Simon Hands.

\appendix*
\section{The lattice Fermi surface}

The conclusions drawn from Figs.~\ref{fig:finitev} and
\ref{fig:lt_12} (i.e. that $\lt<q q_+\rt>$ is approximately
independent of spatial volume and that one would need to use
$L_t\gg20$ to reproduce the curves in Fig.~\ref{fig:qqvsj} without
discarding any data) might suggest that one could successfully reach
the zero 
temperature limit quite cheaply by performing simulations with $L_t\gg L_s$.
For this reason it is worth making the cautionary point that performing
simulations with $\mu\neq0$ on such lattices can lead to unexpected 
results. At low temperature, the Fermi-Dirac distribution resembles a
step function, with the discontinuity about the Fermi momentum smeared
out across a region $\delta k\sim T\sim L_t^{-1}$; in a finite
system, the Brillouin zone is discretised into a cubic momentum
lattice with lattice spacing $2\pi/(a L_s)$. For lattices  with $L_t\gg L_s$,
therefore,
the smearing of the Fermi surface is too fine to be resolved on the
coarse momentum lattice. One consequence is that when the chemical
potential is increased smoothly, the
Fermi-Dirac distribution changes 
only when the Fermi surface crosses a momentum mode. If one were to
study a transition, therefore, such as the chiral crossover
illustrated in Fig.~\ref{fig:EofS}, the physics of the system would be
approximately constant except at points where the surface
crosses a mode;  the transition would be turned into a series of steps.
\begin{figure}[h]
\centering 
\includegraphics[width=11cm]{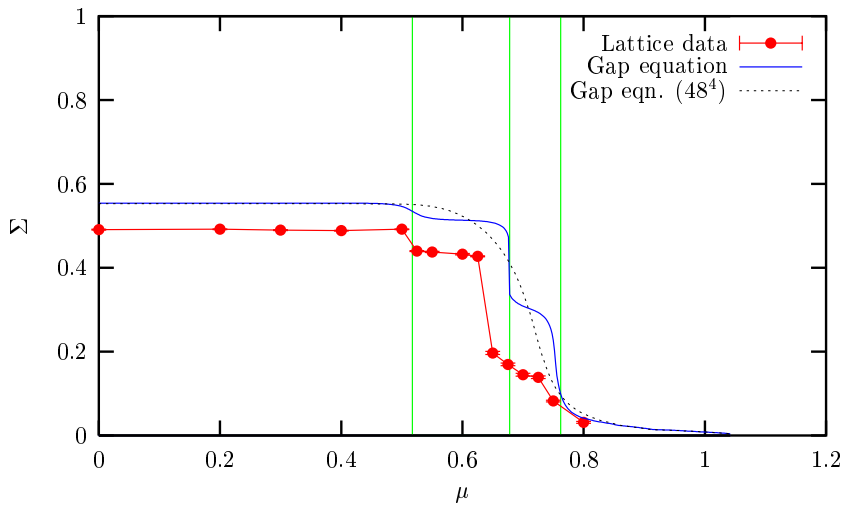}
\caption{Expectation value of the scalar field vs. $\mu$ on a
$10^3\times48$ lattice.}
\label{fig:sigvsmu}
\end{figure}
This is nicely illustrated in Fig.~\ref{fig:sigvsmu}, where the
expectation value of the scalar field $\Sigma\equiv\lt<\sigma\rt>$ is
plotted as a function of $\mu$ on a $10^3\times48$ lattice. The Solid
curve is $\Sigma$ solved via the gap equation in the
large-$N_c$ limit and can be seen to change significantly only when the
solutions of the free-Fermion dispersion relation in the infinite
volume limit
\be
\mu\sim E_F=\sqrt{\sum_{i=1}^3\sin^2{k_F}_i+\Sigma^2}
\ee
coincide with one of the lattice momenta. The first three of these
points are denoted by the vertical lines. The lattice data agree
qualitatively with this curve and the discontinuities 
can be seen clearly. For this reason one cannot make the ratio
$L_t/L_s$ arbitrarily large.

\bibliography{argonne}

\end{document}